# W and Z Production in pp Collisions at 7 TeV with the ATLAS Experiment at the LHC

*V. I. Martinez Outschoorn[1]*
*On behalf of the ATLAS Collaboration*

(1) Harvard University, 17 Oxford St. Cambridge MA 02138, vimartin@physics.harvard.edu

**Abstract**

The observation of *W* and *Z* bosons and a measurement of the production cross sections in proton proton collisions at $E_{CM}$ = 7 TeV are presented using data from the ATLAS experiment at the LHC. Results are based on 118 $W \to l\nu$ and 125 $Z/\gamma^* \to ll$ ($l=e,\mu$) candidate events corresponding to an integrated luminosity of about 17 nb$^{-1}$ and 225 nb$^{-1}$ respectively. The measured values of $\sigma_W \times BR(W \to l\nu)$ = 9.3 ± 0.9 (stat) ± 0.6 (syst) ± 1.0 (lumi) nb and $\sigma_{Z/\gamma^*} \times BR(Z/\gamma^* \to ll, 66 \leq m_{ll} \leq 116)$ = 0.83 ± 0.07 (stat) ± 0.06 (syst) ± 0.09 (lumi) nb. A measurement of the *W* lepton charge asymmetry is also reported. A comparison with theoretical predictions based on NNLO QCD calculations shows agreement with the measurements.

## Introduction

*W* and *Z* bosons decaying to electrons and muons are the first physics signatures with high transverse momentum leptons observed in ATLAS. These processes allow for detailed studies of the detector performance, including lepton reconstruction and trigger efficiencies, momentum scale and resolution. Precise measurements of *W* and *Z* production properties probe perturbative QCD, testing the validity of theoretical NNLO predictions, and provide constraints on PDFs.

The studies are performed using ATLAS [1], a general-purpose detector at the Large Hadron Collider (LHC) designed to provide precision measurements of many objects such as electrons, muons and missing transverse energy. The data are compared to Monte Carlo simulations of signal and background processes using PYTHIA [2] and a detailed ATLAS detector response simulation based on GEANT4 [3]. Details of the *W* and *Z* analyses described here can be found in [4] and [5].

## Event and Lepton Selection

The data were acquired during the first LHC runs at $E_{CM}$ = 7 TeV from March to July 2010, corresponding to an integrated luminosity of about 17 nb$^{-1}$ for the *W* analyses and about 225 nb$^{-1}$ for the *Z*. The uncertainty on the luminosity is estimated to be 11%.

Events were triggered with hardware-based triggers using calorimeter information in $|\eta|<2.5$ for the electron and hit patterns in $|\eta|<2.4$ for the muon. The trigger selection efficiencies are about 100% for single electrons and about 88% for single muons (fully dominated by the trigger detector acceptance). For the muon analyses, scale factors (0.97 ± 0.04 for the *W* dataset and 0.98 ± 0.02 for the *Z*) are used to account for differences in the trigger efficiency between data and Monte Carlo. A primary vertex (PV) with at least three tracks serves to select collision candidates.

Electron identification is based on a sliding-window algorithm forming calorimeter clusters that are then associated to tracks in the Inner Detector (ID). The electron cluster is required to have transverse energy $E_T$ > 20 GeV and to lie within the ID acceptance $|\eta|<2.47$. Clusters in the transition region between the barrel and endcap calorimeters 1.37<|$\eta$|<1.52 are excluded. Different levels of identification criteria are applied to reject backgrounds, including shower shape and hadronic leakage requirements to remove jets, Silicon hit checks to remove conversions, impact parameter cuts to remove heavy-quark decays and searches for secondary maxima in the calorimeter to remove neutral pion decays.

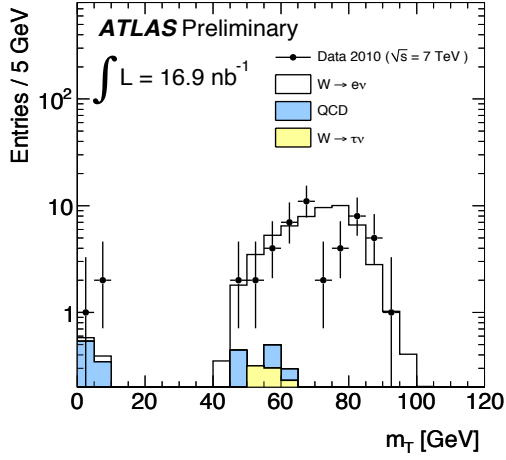

*Fig. 1:* *The $m_T$ distribution for the electron-$E_T^{miss}$ system with $E_T^{miss}$ > 25 GeV. W candidates are selected with $m_T$ > 40 GeV.*

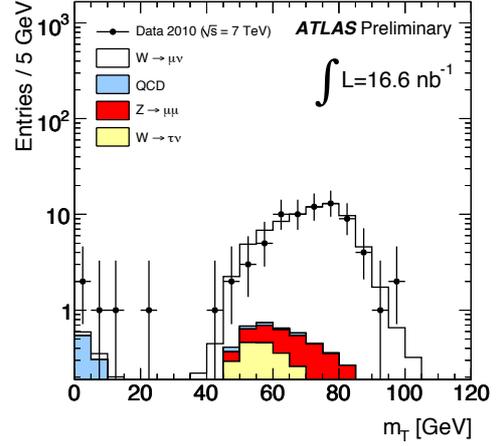

*Fig. 2:* *The $m_T$ distribution for the muon-$E_T^{miss}$ system with $E_T^{miss}$ > 25 GeV. W candidates are selected with $m_T$ > 40 GeV.*

Muon identification is based on combined tracking using both the ID and the Muon Spectrometer (MS). A combined track with transverse momentum $p_T$ > 20 GeV is selected within |$\eta$|<2.4. Cosmic ray contamination is reduced with pointing requirements to the PV in the center of the detector. Additional requirements on the $p_T$ of the MS track and the matching of the ID and MS $p_T$ measurements are applied to reject decays in flight, primarily from pions and kaons. Muons from heavy-quark decays are removed with an isolation requirement, $\Sigma p_T$ of ID tracks in a cone of size 0.4 around the muon divided by the muon $p_T$ < 0.2.

Missing transverse energy $E_T^{miss}$ is based on the sum of calorimeter cell energy deposits in topological clusters at the electromagnetic scale corrected for hadron response and losses due to dead material and out of cluster deposits. The muon calorimeter deposit is replaced with the momentum measurement. Additional quality criteria ensure that events with non-collision large calorimeter energy deposits (e.g. noise) are excluded from the $E_T^{miss}$ calculation.

## W and Z Candidate Selection

The *W* observation is based on the selection of events with a single electron or muon and $E_T^{miss}$ > 25 GeV from the neutrino. Candidates are selected from the peak in the transverse mass distribution $m_T$ > 40 GeV (Figs. 1 and 2). The sample size is of 118 *W→lν* candidates, 46 in the electron channel and 72 in the muon channel. The *Z* event selection is based on two electrons or muons of opposite sign, with loosened identification requirements in the electron analysis. *Z/γ\*→ll* candidates are selected in the invariant mass range 66 < $m_{ll}$ < 116 GeV, yielding 125 events, 46 in the electron channel and 79 in the muon channel (Figs. 3 and 4).

The electroweak background contributions to *W→lν* and *Z→ll* are obtained from Monte Carlo simulation scaled to the NNLO cross section. The electroweak *W→eν* background estimate is 1.5 ± 0.0 (stat) ± 0.1 (syst) events, mainly from leptonic *W→τν* decays. The *W→μν* estimate is 4.4 ± 0.0 (stat) ± 0.3 (syst) events, primarily from *W→τν* and *Z→μμ* where one of the muons falls outside the acceptance.

Data-driven methods are used to estimate the QCD background contribution. Electron QCD backgrounds arise from heavy-quark decays, conversions and hadron fakes. A template fit of the calorimeter isolation yields an estimate of 1.1 ± 0.2 (stat) ± 0.4 (syst) events for *W→eν*. The *W→μν* QCD background, mainly from heavy-quark decays, is estimated from extrapolations from control regions in the $E_T^{miss}$ and isolation plane ("ABCD method"), yielding 0.9 ± 0.3 (stat) ± 0.6 (syst) events.

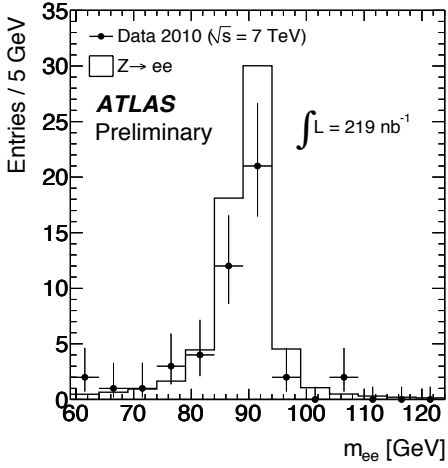

*Fig. 3:* Invariant mass distribution of opposite signed electron pairs.

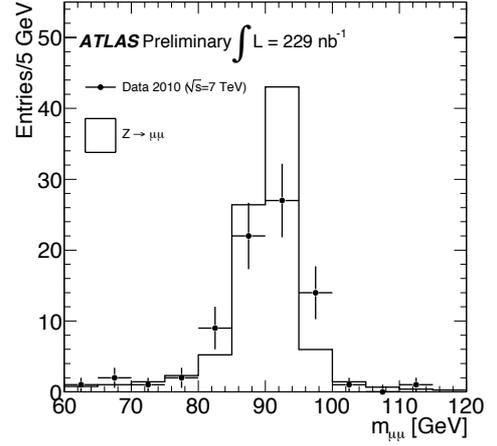

*Fig. 4:* Invariant mass distribution of opposite signed muon pairs.

The estimated backgrounds for the Z are 0.49 ± 0.07 (stat) ± 0.05 (syst) events in the electron channel and 0.17 ± 0.01 (stat) ± 0.01 (syst) events in the muon channel.

## W and Z Cross Section and W Charge Asymmetry

The cross section times branching ratio is

$$\sigma_{W/Z} \times BR = \frac{N_{W/Z}^{sig}}{A_{W/Z} \cdot C_{W/Z} \cdot L_{int}}$$

where $N^{sig}$ is the number of background-subtracted signal events, $A$ is the geometrical acceptance, $C$ includes the trigger and reconstruction efficiencies and $L_{int}$ the integrated luminosity.

In the *W* cross section measurement, the total systematic uncertainty on *C* in the electron channel is 8%, primarily from the lepton identification efficiency. For the muon channel, the total is 7%, mainly from the reconstruction and trigger efficiencies. In the *Z* measurement, the total systematic uncertainty on *C* is 14% for the electron case and 7% for the muon, predominantly from the lepton identification/reconstruction efficiency. The geometrical acceptance is obtained from generator-level Monte Carlo calculations and includes theoretical uncertainties in the modelling of the boson production and from the PDFs. The systematic uncertainty on *A*, based on comparisons with higher order predictions and different PDF sets, is 3% for both *W* and *Z*.

The resulting combined electron and muon cross sections and uncertainties are $\sigma_W \times BR$ $(W \to l\nu)$ = 9.3 ± 0.9 (stat) ± 0.6 (syst) ± 1.0 (lumi) nb where $\sigma_{W+} \times BR(W^+ \to l^+\nu)$ = 5.7 ± 0.7 (stat) ± 0.4 (syst) ± 0.6 (lumi) nb and $\sigma_{W-} \times BR(W^- \to l\nu)$ = 3.5 ± 0.5 (stat) ± 0.2 (syst) ± 0.4 (lumi) nb. The theoretical prediction is calculated with FEWZ [6] using the MSTW2008 NNLO PDF parametrization [7] yielding $\sigma_W \times BR(W \to l\nu)$ = 10.5 ± 0.4 nb, in agreement with the measurement (Fig. 5). The measured Z cross section is $\sigma_Z \times BR$ $(Z/\gamma^* \to ll, 66 \le m_{ll} \le 116)$ = 0.83 ± 0.07 (stat) ± 0.06 (syst) ± 0.09 (lumi) nb in agreement with the theoretical NNLO prediction $\sigma_Z \times BR$ $(Z/\gamma^* \to ll, 66 \le m_{ll} \le 116)$ = 0.96 ± 0.04 nb (Fig. 7).

The measurement of the excess of $W^+$ over $W^-$ in proton proton collisions provides important information about PDFs, particularly the difference in *u* and *d* valence quark contributions. The asymmetry is measured as

$$A_l = \frac{\sigma(l^+) - \sigma(l^-)}{\sigma(l^+) + \sigma(l^-)}, \quad \sigma(l^\pm) = \frac{N_{W^\pm}^{sig}}{C_{W^\pm}}$$

where $\sigma(l^\pm)$ are the fiducial cross sections, satisfying the geometrical and kinematic selection at truth level (in a lepton $|\eta|$ range, $p_T$ > 20 GeV, $E_T^{miss}$ > 25 GeV, $m_T$ > 40 GeV).

For the electron channel, A($|\eta|$<1.37 and 1.52<$|\eta|$<2.47) = 0.21 ± 0.18 (stat) ± 0.01 (syst) and for the muon channel, A($|\eta|$<2.4) =

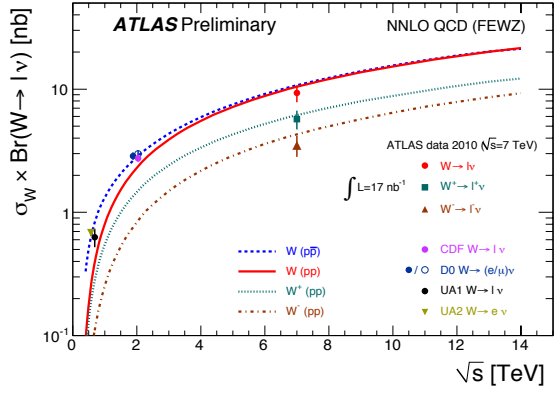

*Fig. 5:* *The measured value of $\sigma_W \times BR(W \to l\nu)$ for $W^+$, $W^-$ and the sum for the combination of electron and muon channels compared to theoretical predictions based on NNLO QCD calculations. The predictions include both proton proton and proton anti-proton collisions as a function of the center-of-mass energy $\sqrt{s}$ and measurements from previous proton anti-proton colliders are also shown.*

0.33 ± 0.12 (stat) ± 0.01 (syst) in agreement with theoretical expectations. Figure 6 shows the asymmetry in two regions of lepton pseudo-rapidity compared to theoretical predictions based on MCatNLO [8] and DYNNLO [9] with PDF sets MSTW2008 [7], CTEQ6.6 [10] and HERAPDF 1.0 [11].

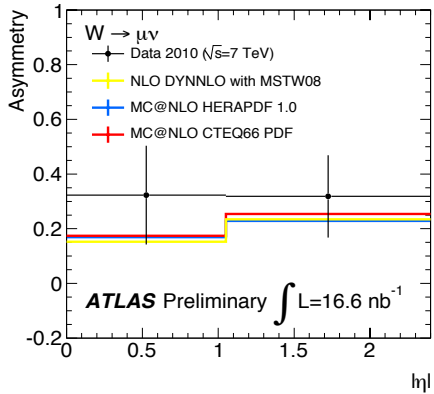

*Fig. 6:* *W charge asymmetry in the muon channel as a function of the lepton pseudo-rapidity compared to various theoretical predictions. The measurement in the electron channel (not shown) is similar.*

## Conclusions

The observation of $W$ and $Z$ bosons decaying to electrons and muons produced in proton proton collisions at $E_{CM} = 7$ TeV are reported. A measurement of the $W$ and $Z$ cross sections and the $W$ charge asymmetry using a dataset corresponding to an integrated luminosity of about 17 nb$^{-1}$ for the $W$ analysis and about 225 nb$^{-1}$ for the $Z$ analysis are presented, in agreement with theoretical predictions based on NNLO QCD.

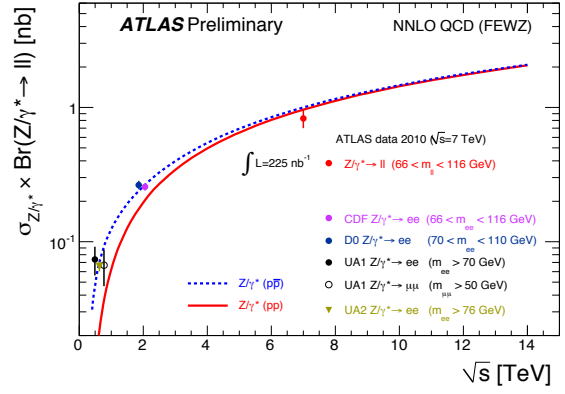

*Fig. 7:* *The measured value of $\sigma_{Z/\gamma^*} \times BR(Z/\gamma^* \to ll)$ for the combination of electron and muon channels compared to theoretical predictions based on NNLO QCD calculations.*

## References

[1] ATLAS Collaboration (2008): *The ATLAS Experiment at the CERN Large Hadron Collider, JINST 3 S08003.*
[2] T. Sjostrand et al. (2006): *PYTHIA 6.4 Physics and Manual, JHEP 05, 026.*
[3] S. Agostinelli et al. (2003): *GEANT4 A Simulation Toolkit, Nucl. Instrum. Methods A 506, 250.*
[4] ATLAS Collaboration (2010): *Measurement of the W->lnu production cross-section and observation of Z->ll production in p-p collisions at sqrt(s) = 7 TeV with the ATLAS detector, ATLAS-CONF-2010-051.*
[5] ATLAS Collaboration (2010): *Measurement of the Z->ll production cross-section in p-p collisions at sqrt(s) = 7 TeV with the ATLAS detector, ATLAS-CONF-2010-076.*
[6] C. Anastasiou et al. (2004): *High-precision QCD at hadron colliders: electroweak gauge boson rapidity distributions at NNLO, Phys. Rev. D69, 094008.*
[7] A.D. Martin et al. (2009):*Parton distributions for the LHC, Eur. Phys. J. C63, 189.*
[8] S. Frixione et al. (2002): *Matching NLO QCD computations and parton shower simulations, JHEP 0206. MCatNLO v3.41 including DY patch.*
[9] S. Catani et al. (2007): *An NNLO subtraction formalism in hadron collisions and its application to Higgs boson production at the LHC, Phys. Rev. Lett. 98.*
[10] P.M. Nadolsky et al. (2008): *Implications of CTEQ global analysis for collider observables, Phys. Rev. D78, 013004.*
[11] H1 Collaboration (2010): *Combined Measurement and QCD Analysis of the Inclusive ep Scattering Cross Sections at HERA, JEP, 01, 109.*